\documentclass[12pt]{iopart}
\usepackage{amssymb}

\newcommand{\btheta}{\mbox{\boldmath $\theta$}}

\def\be{\begin{equation}}
\def\ee{\end{equation}}

\begin{document}

\title{Bayesian Methods for Cosmological Parameter Estimation from Cosmic Microwave
Background Measurements}
\author{Nelson Christensen$^1$, Renate Meyer$^2$, Lloyd Knox$^3$, Ben Luey$^1$ \cite
{email}}

\date{\today}

\address{\
$^{1}$Physics and Astronomy, Carleton College, Northfield, MN 55057,USA
\
}
\address{\
$^{2}$Department of Statistics, The University of Auckland, Auckland, New
Zealand
\
}
\address{\
$^{3}$Department of Physics, University of California - Davis, Davis, CA
95616, USA
\
}

\begin{abstract}
We present a strategy for a statistically rigorous Bayesian approach to the
problem of determining cosmological parameters from the results of
observations of anisotropies in the cosmic microwave background. Our
strategy relies on Markov chain Monte Carlo methods, specifically the
Metropolis-Hastings algorithm, to perform the necessary high--dimensional
integrals. We describe the Metropolis--Hastings algorithm in detail and
discuss the results of our test on simulated data.
\end{abstract}




\section{Introduction}

Recent determinations of the angular power spectrum of the anisotropy
of the cosmic microwave background (CMB) \cite
{cbi01,max00,boom1,TOCO2,mauskopf00,PythV,Viper} have created much
excitement. These data may be used to estimate cosmological
parameters. The maximum in the angular power spectrum around the
multipole value $l\thickapprox 200$ ('the first peak') is consistent
with inflation and vanishing mean spatial curvature
\cite{boom3,Maxima1,Knox}.  The lack of curvature adds compelling
evidence to the case for the existence of some smoothly--distributed
negative--pressure component which is possibly a cosmological constant
(e.g., \cite{dodknox00}). The constraints on the amplitude of the
predicted second peak lead to lower limits on the baryon density
\cite{boom3,Jaffe00,Max} which are in conflict with the standard
results from big bang nucleosynthesis (e.g., \cite{burles00}); new
data are eagerly awaited to provide a resolution. Fortunately, future
satellite missions will provide angular power spectrum determinations
with greatly improved precision \cite{MAP,Planck}.

The extraction of cosmological parameters from the CMB\
anisotropy data requires a long and computer--intensive analysis
chain.  Here we are interested in the last step in the chain which is
a derivation of cosmological parameters from the inferred angular
power spectrum, $C_l$.

This last step is computer--intensive as well and promises to be more
so as the number of parameters in ones model increases.  Current
analysis exercises have included up to ten 
parameters\cite{boom3,Jaffe00,Max}; the scalar
quadrupole and gravity wave perturbation normalizations ($A_{s}$ and
$A_{t}$ ), the scalar and tensor power--law indices for primordial
perturbations ($ n_{s}$ and $n_{t}$), the reionization optical depth
$\tau $, the spatial curvature $\Omega _{k}$, the energy densities for
baryonic matter ($\omega _{b}$), cold dark matter ($\omega _{cdm}$),
neutrinos ($\Omega _{\nu }$) and the vacuum ($\Omega _{\Lambda }$).  
Other logical cosmological parameters could include the
number of neutrino families and their masses.  Potentially infinite
degrees of freedom lie in the primordial spectrum of perturbations.

The estimation of parameters can be thought of as an exercise in Bayesian
inference. One starts with the likelihood function, namely the conditional
probability distribution function (PDF) of the observation $\mathbf{z}$ 
\emph{given} the unknown parameters, $p(\mathbf{z}|\mbox{\boldmath
$\theta$})$. For studies pertaining to the CMB, the likelihood used is an
approximation to the probability of the power spectrum given the data, e.g. 
\cite{BJK00}. If the likelihood function for a certain parameter vector $
\mbox{\boldmath $\theta$}_{1}$ is higher than for a parameter vector $
\mbox{\boldmath $\theta$}_{2}$, the observations are more likely to have
occurred under $\mbox{\boldmath
$\theta$}_{1}$ than under $\mbox{\boldmath $\theta$}_{2}$. Thus, the
observations give higher plausibility to $\mbox{\boldmath $\theta$}_{1}$ and
finding the parameter value that maximizes the likelihood seems to have some
compelling logic. However, the only consistent way to quantify one's
uncertainty about an \emph{unknown} parameter $\mbox{\boldmath $\theta$}$ is
by specifying a probability distribution or equivalently a PDF for the
parameter. After observing the data, this distribution should be updated to
the PDF of the parameter \emph{given} the data by incorporating the gained
information. We are not interested in the PDF of the data given the \emph{
unknown} $\mbox{\boldmath $\theta$}$ but in the PDF of $
\mbox{\boldmath
$\theta$}$ given the \emph{known} data; we want 
the ``inverse probability'', not $p(
\mathbf{z}|\mbox{\boldmath $\theta$})$ but $p(\mbox{\boldmath $\theta$}|
\mathbf{z})$. The only coherent approach to update a \emph{prior}
probability distribution with experimental information consists of
calculating the \emph{posterior} PDF via Bayes' theorem: 
\begin{equation}
p(\mbox{\boldmath $\theta$}|\mathbf{z})=\frac{p(\mbox{\boldmath $\theta$})p(
\mathbf{z}|\mbox{\boldmath $\theta$})}{m(\mathbf{z})}
\end{equation}
where $m(\mathbf{z})=\int p(\mathbf{z}|\mbox{\boldmath $\theta$})f(
\mbox{\boldmath $\theta$})d\mbox{\boldmath $\theta$}$ is the marginal PDF of 
$\mathbf{z}$. The denominator 
can be regarded as a normalization constant since it is
independent of $\mbox{\boldmath $\theta$}$. The posterior PDF thus combines
prior with likelihood information. A Bayesian point estimate of the unknown
parameter would be the posterior mean, median, or mode.

The usual difficulties of Bayesian inference apply here: namely, the
challenge of high--dimensional integration. One needs to be able to perform
high-dimensional integrations when calculating the normalization constant $m(
\mathbf{z})$, for instance, or when calculating the marginal PDF of a single
component of the parameter vector $\mbox{\boldmath $\theta$}$ by integrating
out all the other components, or when calculating posterior means. One of
the main objectives of this paper is to demonstrate the potential of
what statisticians refer to as simulation--based integration techniques
for Bayesian posterior computation of CMB model parameters.  These
are called simulation--based, not because there is any simulation
of the data, but because the posterior distribution is simulated.
The particular simulation--based method we use here generates
a set of parameter vectors whose distribution simulates the posterior
distribution. 

Analyses of cosmological parameter constraints from the most recent angular
power spectrum determinations start with the calculation of the likelihood
on an $n_{p}$--dimensional grid, where $n_{p}$ is the number of parameters
\cite{boom3,Jaffe00,Max}.
The next step is then the reduction of this large amount of data to a series
of one-- or two--dimensional probability distributions functions which can
easily be plotted. This is achieved by marginalizing over the other
variables; i.e., integrating over them. Note that some times (and in all
applications for at least some of the variables) this marginalization is
approximated by a \textit{maximization} over the other variables.

The chief drawback to the grid--based approach is the exponential increase
in computing times and storage requirements with increasing number of
parameters.  Even with $n_p = 10$ (plus a handful of instrumental parameters)
overly coarse grids and approximate treatment of some of the
marginalizations are necessary to make the problem tractable. Including
more parameters, for example ones that describe instrumental effects,
foreground contaminants, or departures of the primordial power spectrum from
a strict power--law, will make the grid--based approaches even more
difficult if not impossible.

In this paper we implement the Bayesian approach to estimation of
cosmological parameters using computer--intensive Markov chain Monte Carlo
(MCMC) methods.  Impediments to multidimensional integrations have been
overcome by the progress made within the last decade in Bayesian
computational technology via MCMC methods \cite{gilk96}. Since its initial
application in digital signal analysis \cite{Gem} MCMC\ methods have
revolutionized many areas of applied statistics and we expect there to be an
impact on cosmological parameter estimation from CMB measurements as well. A
distinct advantage of the MCMC\ approach is that computational time does not
grow exponentially with parameter number, as it does for other methods \cite
{gilk96}. MCMC\ techniques have been applied in numerous areas, from science
to economics \cite{meye00}. Applications of state--space modeling in finance,
e.g. stochastic volatility models applied to time series of daily exchange
rates or returns of stock exchange indices, easily have 1000--5000 parameters 
\cite{jacq94}. Specially tailored MCMC algorithms can markedly improve the
calculational speed \cite{meyu00,kim98}.\ Hence, the MCMC approach to
cosmological parameter estimation may provide the best strategy when testing
complex models with numerous parameters. 

We test our MCMC parameter estimation routine using simulated data. The
likelihood is produced via a prototype fast calculator\cite{knoxskordis01}.
The toy model depends on four parameters.  We demonstrate the validity of the
technique with this example. The goal of our research is to apply MCMC\
methods to real CMB data, and we are currently optimizing our routines for
this multiparameter (\symbol{126}10) analysis. We are optimistic since
experience has generally shown scaling of computing times with the number of
parameters to be slower than exponential \cite{gilk96}.

In section 2 we review methods that have been used to estimate cosmological
parameters from CMB\ measurements, plus techniques for calculating
likelihoods. In section 3 we describe the Bayesian approach to statistical
inference and its implementation via Markov chain Monte Carlo methods. In
section 4 we describe a method for applying MCMC\ methods to cosmological
parameter estimation with CMB\ data. We test our method with a four
parameter example, and our results are presented in Section 5. Section 6
presents our conclusions.


\section{Current Statistical Methods with CMB\ Data}


\subsection{Parameter Estimation}

Parameter estimation from CMB data is usually performed with a multi--step
process, the last two steps of which are determining the power--spectrum
from a map (or otherwise--pixelized data) \cite{Tegmark97,osh,BJK98}, 
and then determining the
cosmological parameters from the power spectrum. To date it has been
impractical to go straight from the map to the parameters because of the
computational expense of evaluating the probability of the map given the
parameters thousands of times. Fortunately, the structure of the probability
of the map given the power spectrum (the likelihood of the power spectrum)
is much simpler than that of the map given the parameters.  The simple
structure allows for the mode to be found in a small number of iterations
of a Levenberg--Marquardt--type search algorithm \cite{BJK98}.

This ``radical'' compression of the information in the map to information in
the power spectrum is possible because we assume that the signal in the maps
is Gaussian and statistically isotropic. However, the uncertainties in the
resulting power spectrum are not normally--distributed and the above search
procedures do not allow one to completely characterize the distribution.
Fortunately there are analytic approximations to the complete distribution
whose parameters one can calculate with minor adaptations of the
power--spectrum mode search algorithms \cite{BJK00,Bart1}.

The most recent attempts at determining cosmological parameters
(\cite{boom3,Jaffe00,Max}) are in fact
attempts to determine the posterior probability distribution of the
cosmological parameters under various prior assumptions. The first step is
to evaluate the likelihood on a grid of cosmological parameters. To have
constraints of reduced--dimensionality (one or two) suitable for plotting,
one marginalizes over the other parameters. Sometimes marginalization is
approximated by simply maximizing over the remaining parameters. This
approximation is exact if the marginalized parameters are normally
distributed.

Direct grid--based evaluations of the likelihood have computation time and
storage requirements which rise exponentially with the number of parameters.
Such approaches will be difficult to implement for models of greater
complexity than have been studied so far. Further, the maximization
approximation to marginalization can sometimes lead to spurious results.
Numerical maximization techniques, such as the Levenberg--Marquardt method,
are only guaranteed to find a local maximum. Once they have reached a local
maximum they might get stuck in their search and not reach the global
maximum. For this reason statisticians have applied \emph{simulated
annealing}, which is a technique for global optimization. Simulated
annealing has been attempted for cosmological parameter estimation \cite
{Hann,Knox3}. Simulated annealing is related to MCMC as its core component
is the Metropolis Hastings algorithm. In this method the parameter space is
searched in a random way. A new parameter space point is reached with a
probability that depends on the likelihood and an \textit{effective
temperature} term. In the limit where the temperature approaches zero the 
\textit{thermodynamics} of this parameter space search finds the system
approaching the maximum of the likelihood. Although these methods are
applicable to high--dimensional problems, they can be very inefficient in
certain situations. Furthermore, there is no guarantee that the global
maximum will be reached in a finite time. The efficiency depends very much
on specifying a good cooling schedule which involves the arbitrary and
skillful choice of various cooling parameters.

\subsection{Calculating the Likelihood}

Implicit in the problem of parameter estimation, whether from a frequentist
or Bayesian perspective, is the calculation of the likelihood of the
observed power spectrum (determined from a map) given some cosmological
model. Thus likelihood calculation requires us to be able to calculate the
angular power spectrum for a given model (with its associated cosmological
parameters). This calculational task is accomplished with computer
codes such as CMBfast 
\cite{CMBFAST} or CAMB \cite{CAMB}. These codes accept the cosmological
parameters as input, and return the angular power spectrum of the CMB\
anisotropies, $C_{l}$. These software packages serve as the work--horses of
current CMB\ parameter--determination 
efforts. For example, the likelihood has been
calculated 30,311,820 times in order to cover a region in a ten--dimensional
cosmological parameter space \cite{Max}. In other studies the likelihood was
evaluated as needed within the calculation \cite{dodknox00,boom2}.

\section{Bayesian Posterior Computation via MCMC}

Parameter estimation can be comprehensively described within the language of
Bayesian inference. Application of Bayes' theorem is well--suited to
astrophysical observations \cite{Lor}. In Bayesian data analysis the model
consists of a joint distribution over all unobserved (parameters) and
observed (data) quantities. One conditions on the data to obtain the
posterior distribution of the parameters. The starting point of the Bayesian
approach to statistical inference is setting up a full probability model
that consists of the \emph{joint} probability distribution of all
observables, denoted by $\mathbf{z}=(z_{1},\ldots ,z_{n})$ and unobservable
quantities, denoted by $\mbox{\boldmath $\theta$}=(\theta _{1},\ldots
,\theta _{d})$. Using the notion of conditional probability, this joint PDF, 
$p(\mathbf{z},\mbox{\boldmath $\theta$})$can be decomposed into the product
of the PDF of all unobservables, $p(\mbox{\boldmath $\theta$})$, referred to
as the \emph{prior}\/ PDF of $\mbox{\boldmath $\theta$}$, and the conditional
PDF of the observables given the unobservables, $p(\mathbf{z}|
\mbox{\boldmath
$\theta$})$, referred to as the sampling distribution or \emph{likelihood},
i.e.\ 
\begin{equation}
p(\mathbf{z},\mbox{\boldmath $\theta$})=p(\mbox{\boldmath $\theta$})p(
\mathbf{z}|\mbox{\boldmath $\theta$}).
\end{equation}
The prior PDF contains all the information about the unobservables that is
known from substantive knowledge and expert opinion \emph{before} observing
the data. All the information about the $\mbox{\boldmath $\theta$}$ that
stems from the experiment is contained in the likelihood. In the light of
the data, the Bayesian paradigm then updates the prior knowledge about $
\mbox{\boldmath $\theta$}$, $p(\mbox{\boldmath $\theta$})$, to the \emph{
posterior} PDF of $\mbox{\boldmath $\theta$}$, $p(\mbox{\boldmath $\theta$}|
\mathbf{z})$. This is done via an application of Bayes' theorem through
conditioning on the observations 
\begin{equation}
p(\mbox{\boldmath $\theta$}|\mathbf{z})=\frac{p(\mbox{\boldmath $\theta$},
\mathbf{z})}{m(\mathbf{z})}\propto p(\mbox{\boldmath $\theta$})p(\mathbf{z}|
\mbox{\boldmath
$\theta$})
\end{equation}
where $m(\mathbf{z})=\int p(\mathbf{z}|\mbox{\boldmath $\theta$})p(
\mbox{\boldmath
$\theta$})d\mbox{\boldmath $\theta$}$ is the marginal PDF of $\mathbf{z}$
which can be regarded as a normalizing constant as it is independent of $
\mbox{\boldmath $\theta$}$. The Bayesian approach is based on the likelihood
function but also quantifies the uncertainty about  the parameters
 through a joint
prior distribution that summarizes the available information about the
parameters before observing the data. In the light of the observations, the
information about the unknown parameters is then updated via Bayes' theorem
to the posterior distribution which is proportional to the product of
likelihood and prior density \cite{carl96}.

As already mentioned in the introduction, the main difficulty with the
Bayesian approach to parameter estimation is high--dimensional integration.
To calculate the normalizing constant of the joint posterior PDF, for
instance, requires $d$--dimensional integration. Having obtained the joint
posterior PDF of $\mbox{\boldmath $\theta$}$, the posterior PDF of a single
parameter $\theta _{i}$ of interest can be obtained by integrating out all
the other components, i.e. 
\begin{equation}
p(\theta _{i}|\mathbf{z})=\int \ldots \int p(\mbox{\boldmath $\theta$}|
\mathbf{z})d\theta _{1}\ldots d\theta _{i-1}d\theta _{i+1}\ldots d\theta
_{d}.
\end{equation}
Calculation of the posterior mean of $\theta _{i}$ necessitates a further
integration, e.g.\ $E[\theta _{i}|\mathbf{z}]=\int \theta _{i}p(\theta _{i}|
\mathbf{z})d\theta _{i}$.  This procedure is referred to as marginalization.

Only in the simplest situations can these integrals be solved
analytically. The main approximate techniques are normal and Laplace
approximations based on asymptotics, quadrature approximations, Monte Carlo
integrations, and stochastic simulation. MCMC methods belong to the last
category. For an overview see \cite{Evans95}. The deterministic techniques
rely upon approximate normality and asymptotic results in the sense of the
sample size growing to infinity. These techniques were mostly developed
before the immense technological advances that enabled computer--intensive
methods to be applied. The complexity of these techniques increases
substantially with the dimension of the parameter space. In very broad
terms, experience suggests that deterministic techniques provide good
results for low--dimensional models. Similar comments are valid for
non--iterative simulation techniques, since finding a suitable auxiliary
distribution in rejections or importance sampling, for instance, becomes an
extremely difficult task for high dimensions. As the dimension of the model
increases, only iterative simulation--based integration
techniques such as sampling--importance
resampling (SIR) or MCMC provide adequate
solutions. One major advantage of using a sampling--based approach to
posterior computation is that once a sample from the posterior PDF, say $(
\mbox{\boldmath $\theta$}_{1},\ldots ,\mbox{\boldmath $\theta$}_{N})$ is
available, we can employ this to estimate the posterior mean of each
parameter by the sample average of the corresponding component, the marginal
PDF's using kernel density estimates, and correlation between parameters
using the sample correlations.

To avoid a time--consuming direct sampling of the joint posterior 
we propose a MCMC method \cite{gilk96,chri98}. 
Instead of generating a sequence of
independent samples from the joint posterior, in MCMC a Markov chain is
constructed whose equilibrium distribution is just the joint posterior.
Thus, after running the Markov chain for a certain \textit{burn--in} period,
one obtains (correlated) samples from the limiting distribution, provided
that the Markov chain has reached convergence.

One method for constructing a Markov chain is via the Metropolis--Hastings
algorithm. The MH algorithm was developed by Metropolis \textit{et al.}\ 
\cite{metr53} and generalized by Hastings \cite{Hast}. It is a MCMC method
which means that it generates a Markov chain whose equilibrium distribution
is just the target density, here the joint posterior PDF $p(\mbox{\boldmath
$\theta$}|\mathbf{z})$. This means that after sampling from this Markov
chain for a sufficiently long time to allow the chain to reach equilibrium,
the samples can be regarded as samples from the joint posterior PDF.

The MH algorithm shares the concept of a generating PDF with the well--known
simulation technique of \emph{rejection sampling}, where a candidate is
generated from an auxiliary PDF and then accepted or rejected with some
probability. However, the candidate generating PDF, $q(
\mbox{\boldmath
$\theta$}|\mbox{\boldmath $\theta$}_{n})$ can now depend on the current
state $\mbox{\boldmath $\theta$}_{n}$ of the Markov chain. A new candidate $
\mbox{\boldmath $\theta$}^{\prime }$ is accepted with a certain \emph{
acceptance probability} $\alpha (\mbox{\boldmath $\theta$}^{\prime }|
\mbox{\boldmath
$\theta$}_{n})$ also depending on the current state $
\mbox{\boldmath
$\theta$}_{n}$ given by: 
\begin{equation}
\alpha (\mbox{\boldmath $\theta$}^{\prime }|\mbox{\boldmath $\theta$}
_{n})=\min \left\{ \frac{p(\mbox{\boldmath $\theta$}^{\prime })p(\mathbf{z}|
\mbox{\boldmath $\theta$}^{\prime })q(\mbox{\boldmath $\theta$}_{n}|
\mbox{\boldmath $\theta$}^{\prime })}{p(\mbox{\boldmath $\theta$}_{n})p(
\mathbf{z}|\mbox{\boldmath $\theta$}_{n})q(\mbox{\boldmath $\theta$}^{\prime
}|\mbox{\boldmath $\theta$}_{n})},1\right\}
\end{equation}
if $(p(\mbox{\boldmath $\theta$}_{n})p(\mathbf{z}|\mbox{\boldmath $\theta$}
_{n})q(\mbox{\boldmath $\theta$}^{\prime }|\mbox{\boldmath $\theta$}_{n}))>0$
and $\alpha (\mbox{\boldmath
$\theta$}^{\prime }|\mbox{\boldmath $\theta$}_{n})=1$ otherwise. \bigskip
The steps of the MH algorithm are therefore: \bigskip

\begin{tabular}{ll}
Step 0: & Start with an arbitrary value $\mbox{\boldmath
$\theta$}_{0}$ \\ 
&  \\ 
Step $n+1$: & Generate $\mbox{\boldmath
$\theta$}^{\prime }$ from $q(\mbox{\boldmath
$\theta$}|\mbox{\boldmath $\theta$}_{n})$ and $u$ from $U(0,1)$ \\ 
& If $u\leq \alpha (\mbox{\boldmath $\theta$}^{\prime }|
\mbox{\boldmath
$\theta$}_{n})$ set $\mbox{\boldmath $\theta$}_{n+1}=
\mbox{\boldmath
$\theta$}^{\prime }$ (acceptance) \\ 
& If $u>\alpha (\mbox{\boldmath $\theta$}^{\prime }|\mbox{\boldmath $\theta$}
_{n})$ set $\mbox{\boldmath $\theta$}_{n+1}=\mbox{\boldmath $\theta$}_{n}$
(rejection)
\end{tabular}
\bigskip

The MH algorithm does not require the normalization constant of the target
density. Note that the outcomes from the MH algorithm can be regarded as a
sample from the invariant density only after a certain \textit{burn--in}
period. Although the theory guarantees convergence for a wide variety of
proposal PDFs, it does not say anything about the speed of convergence, i.e.
how long the ''burn--in'' period should be. Convergence rates of MCMC
algorithms are important topics of ongoing statistical research with little
practical findings so far. There is no formula for determining the minimum
length of an MCMC run beforehand, nor a method to confirm that a given chain
has reached convergence. The only tests available are based on an empirical
time series analysis of the sampled values and can only detect
non--convergence. Thus, by performing a whole sequence of so--called
convergence diagnostics with negative results, one only gains more
confidence in one's hope that the chain reached equilibrium but never a
guarantee. For issues concerning convergence diagnostics, the reader is
referred to Cowles and Carlin \cite{CC}.

Various convergence diagnostics have been developed and are implemented in
CODA \cite{coda95}. CODA is a menu--driven
collection of functions for analyzing the output of the Markov chain.
Besides trace plots and the usual tests for convergence, CODA calculates
statistical summaries of the posterior distributions and kernel density
estimates.

\section{Applying MCMC\ Methods to CMB\ Anisotropy Data}

Probability distribution functions for cosmological parameters given CMB
anisotropy data can be computed in a Bayesian fashion with the MCMC serving
as a means of conducting a proper marginalization over parameters. The
implementation of the MCMC method is relatively straightforward. Instead of
calculating the likelihood at uniform locations in the parameter space \cite
{boom3,Jaffe00,Max}, 
one lets the MCMC do its intelligent walk through the space.
Uniform \emph{a priori} distributions for the parameters seem reasonable, so
the MCMC\ could sample the parameter space defined by them. Since the
likelihood function can not be written explicitly in terms of the
cosmological parameters, but instead in terms of the CMB\ anisotropy power
spectra, $C_{l}$, it is necessary to implement a Metropolis--Hastings MCMC
routine.

The Markov chain can commence at a randomly selected position in parameter
space $(\theta _{1}^{(0)},\ldots ,\theta _{d}^{(0)})$. With the parameter
set one can then utilize a program like CMBfast \cite{CMBFAST} or CAMB \cite
{CAMB} to generate an angular power spectrum and a likelihood can then be
calculated. New values for the parameters $(\theta _{1}^{(1)},\ldots ,\theta
_{d}^{(1)})$ are selected via sampling from the \emph{a priori}
distributions. However, these values are not necessarily \emph{accepted} as
new values. First their likelihoods are evaluated. Then the new values are 
\emph{accepted} or rejected according to the following test; a random
number, $u$, would be generated between 0 and 1. If
 \ $u\leq \min \left[ 1,(p(\theta _{1}^{(1)},\ldots ,\theta _{d}^{(1)})p(
\mathbf{z}|\theta _{1}^{(1)},\ldots ,\theta _{d}^{(1)}))/(p(
\theta _{1}^{(0)},\ldots ,\theta _{d}^{(0)})p(\mathbf{z}|\theta
_{1}^{(0)},\ldots ,\theta _{d}^{(0)}))\right] $ where $p(\mathbf{z}|\theta
_{1}^{(1)},\ldots ,\theta _{d}^{(1)})$ is the likelihood in terms of data $
\mathbf{z}$ and cosmological parameters $(\theta _{1},\ldots ,\theta _{d})$
then the new parameters is \emph{accepted} into the chain, if not the next
chain element has values equal to that of the previous state. A new set of
parameters is then randomly sampled from the a priori distributions and the
procedure continues.

The generated chain of parameter values forms the set from which the
statistical properties would be derived. After running the Markov chain for
a certain burn--in period (in order for the Markov chain to reach
convergence) one obtains (correlated) samples from the limiting
distribution. This process continues for a sufficiently long time (as
determined by convergence diagnostics \cite{coda95}).

After the \emph{burn--in} the frequency of appearance of parameters
represents the actual posterior density of the parameter. From the posterior
density one can then create confidence intervals. Summary statistics are
produced from the distribution, such as posterior mean and standard
deviation. A cross--correlation matrix is also easily produced---which could
prove to be a useful part of quantifying the multi--dimensional constraints.

The above method constitutes the simplest implementation of the
Metropolis--Hastings method, that of an independence chain. We have just used
an \emph{acceptance probability} $\alpha (\mbox{\boldmath $\theta$}^{\prime
}|\mbox{\boldmath $\theta$})$, defined by 
\begin{equation}
\alpha (\mbox{\boldmath $\theta$}^{\prime }|\mbox{\boldmath $\theta$})=\min
\left\{ \frac{p(\btheta')p(\mathbf{z}|\mbox{\boldmath $\theta$}^{\prime })q(
\mbox{\boldmath $\theta$}|\mbox{\boldmath $\theta$}^{\prime })}{
p(\btheta)p(\mathbf{z}|
\mbox{\boldmath $\theta$})q(\mbox{\boldmath $\theta$}^{\prime }|
\mbox{\boldmath $\theta$})},1\right\}
\end{equation}
where the generating density $q(\mbox{\boldmath $\theta$}^{\prime }|
\mbox{\boldmath
$\theta$})$ is the uniform density over the parameter space and thus, in
particular, independent of the current state. By using uniform priors, the
posterior PDFs in the acceptance probability calculation reduce to the
likelihoods. However, the efficiency of a Metropolis--Hastings algorithm
depends crucially on the form of the generating density $q(
\mbox{\boldmath
$\theta$}^{\prime }|\mbox{\boldmath $\theta$})$. Just using a uniform
distribution that does not even depend on the current state 
{\boldmath$\theta $} is the
simplest but probably least efficient way to accomplish the task. Even
with a uniform distribution, the algorithm will be
irreducible/aperiodic/reversible and thus the Markov chain will converge
towards its stationary distribution.

A slightly better way might be to use a uniform distribution in a
neighborhood of the current $\mbox{\boldmath $\theta$}$. 
Any prior
information could be useful, such as correlations that one could use to
specify a multivariate normal centered around the current $
\mbox{\boldmath
$\theta$}$ with a covariance matrix that takes said correlations into
account. The optimization of the Metropolis--Hastings MCMC strategy will
inevitably require experimentation with the generating density $q(
\mbox{\boldmath $\theta$}^{\prime }|\mbox{\boldmath $\theta$})$. While this
may require some detailed study, the benefit will be the ability to generate
posterior distributions for a large number of cosmological parameters.

An approximate formula for the Fisher matrix of the cosmological parameters
which has often been used for forecasting parameter errors for CMB
experiments may provide us with a useful generating density. The Fisher
matrix is the expectation value of the second--derivative of the log of the
likelihood and is given by: 
\begin{equation}
\label{eqn:paramfish}
F_{pp^{\prime }}\left( \mbox{\boldmath $\theta$}\right) =\sum_{l}{\frac{
\partial C_{l}}{\partial \theta _{p}}}{\frac{\partial C_{l}}{\partial \theta
_{p^{\prime }}}}{\frac{1}{\sigma _{l}^{2}}}
\end{equation}
where the variance in each $C_{l}$ determination can be approximated by 
\begin{equation}
\label{eqn:sigcl}
\sigma _{l}^{2}={\frac{2}{(2l+1)f_{\mathrm{sky}}}}\left(
C_{l}+w^{-1}B_{l}^{-2}\right) ^{2}
\end{equation}
where $f_{\mathrm{sky}}$ is the fraction of sky observed, $w^{-1}=\sigma _{
\mathrm{pix}}^{2}\Omega _{\mathrm{pix}}$, $\sigma _{\mathrm{pix}}$ is the
standard error in each map pixel, $\Omega _{\mathrm{pix}}$ is the size of
the pixel and $B_{l}$ is the Legendre transform of the beam profile. The
generating density would then be 
\begin{equation}
q(\mbox{\boldmath $\theta$}^{\prime }|\mbox{\boldmath $\theta$})\propto
\exp \left(
-\sum_{pp^{\prime }}(\theta _{p}-\theta _{p}^\prime )F_{pp^{\prime }}\left( 
\mbox{\boldmath $\theta^\ast$}\right) (\theta _{p^{\prime }}-\theta _{p^{\prime
}}^{\prime })/2\right)
\end{equation}
Note that the Fisher matrix is always evaluated at the same 
${\bf \btheta}^\ast$ to avoid having to calculate the derivatives in
Eq. ~\ref{eqn:paramfish} at each iteration.

One may wish to know the posterior distribution of the
parameters for a range of choices of priors.  For example,
in \cite{boom3} the analysis was performed with 13 different
priors each of which corresponded to a different choice about our
knowledge of, e.g., the Hubble constant and the baryon density.
For some parameter--determination methods adoption of a variety
of priors is very costly since for each choice of prior the
entire calculation must be redone.  This is true, for example,
of the Levenberg--Marquardt--type search algorithm used in
\cite{dodknox00}.  In contrast, as emphasized in \cite{Max},
evaluation of the likelihood on a grid allows for rapid calculation
of various posteriors with different prior assumptions.

Fortunately, the sampled values from a Markov chain constructed 
assuming one prior can be used for other priors as well
using {\em importance sampling} \cite{Ripley87}. 
Suppose we have devised a MCMC sampler with stationary distribution
$p^*(\theta|z)\propto p_1(\theta)p(z|\theta)$ with a prior $p_1$.
Then we can estimate the expectation of an
arbitrary function $g(\theta)$ of interest
under the modified posterior
$p(\theta|z)\propto p_2(\theta)p(z|\theta)$, i.e.\
using a new prior $p_2$, by importance reweighting the output
$\theta_1,\ldots,\theta_N$ from the chain with stationary 
distribution $p^*$. Thus,
\[
E_p[g(\theta)|z] \approx \sum_{i=1}^N \frac{w_i g(\theta_i)}{\sum_{i=1}^N
w_i}\]
where the importance weight is $w_i=p_2(\theta_i)/p_1(\theta_i)$.

\section{Test of Metropolis--Hastings Algorithm}

We have successfully tested our MH method on simulated data. The `data' we
simulated is an angular power spectrum from $l=100$ to $l=800$ with normally
distributed errors. The underlying model from which we made the realization
is a cold dark matter model with baryonic matter density $\omega_b = 0.019$,
dark matter density $\omega_d = 0.154$, $A=100.0$ (in some units) and $n=1.0$
. The errors were realized as uncorrelated Gaussian random variables with
zero mean and variance, $\sigma^2_l$ given by Eq.~\ref{eqn:sigcl}
assuming a noise--free map covering 10\% of the sky.

To evaluate the likelihood of the data at any given point in the
four--dimensional parameter space we evaluated: 
\begin{equation}
-2\ln{\mathcal{L}} = \sum_l\left(C_l(\omega_b,\omega_d,A,n) - C_l^{\mathrm{
data}}\right)^2/\sigma_l^2.
\end{equation}
The most time--consuming step in the likelihood calculation is calculation
of $C_l(\omega_b,\omega_d,A,n)$. We sped this up using a fast $C_l$
calculator described very briefly here and in more detail elsewhere \cite
{knoxskordis01}. We use CMBfast to pre--compute $\Delta_l^2(k)$ on an 8 by 8
grid of values of $\omega_b$ and $\omega_d$. The quantity $\Delta_l^2(k)$
gives the contribution from each wave number to the angular power spectrum: $
C_l = (4\pi)^2\int dk k^2 \Delta_l^2(k) P(k)$ where $P(k)$ is the primordial
spectrum of density perturbations which we parametrize as a power--law: $
P(k)=Ak^n$. Thus, for a given $\omega_b,\omega_d,A$ and $n$ we perform a
multi--linear interpolation over the $\omega_b$, $\omega_d$ grid of $
\Delta_l^2(k)$ values and then do the integral over $k$.

A simple Fortran code was used for the MH routine, and it can be obtained at 
\textit{http://physics.carleton.edu/Faculty/nelsonhome.html}. This routine
calls the fast likelihood calculator for the likelihood values. The \textit{
a priori} distributions were all uniform, with ranges of $0.015<\omega
_{b}<0.025$, $0.1<\omega _{d}<0.2$., $0<n<2$ and $0<A<200$. The candidate
generating densities were uniform and centered at the current chain
parameter values and extending with ranges of $5 \times 10^{-4}$ for $\omega _{b}$, 
$5\times 10^{-3}$ for $\omega _{d}$, $0.25$ for $A$, and $0.01$ for $n$. The
results presented here were from a run of 200000 iterations, with an
acceptance ratio of 41\%. This calculation took \symbol{126}24 hours on a
Sun Ultra 10 (440 MHz) workstation.

For our analysis we thinned the 200,000 cycle chain by accepting every 20th
observation in order to avoid highly correlated values. Of the remaining
10,000 samples we used a burn--in of 1,000 which yields a final chain length
of 9,000. Extensive convergence diagnostics were calculated for the four
parameters using the CODA\ software \cite{coda95}. All chains passed the
Heidelberger--Welch stationarity test. The Raferty--Lewis convergence
diagnostics confirmed that the thinning and burn--in period were sufficient.
Lags and autocorrelations within each chain were reasonably low for all
parameters. These convergence diagnostics are summarized in \cite{coda95},
and references therein.

The trace plots and resulting kernel densities for the four parameters are
shown in figure 1. Summary statistics including posterior mean, standard
deviation, the time series standard error (the square root of the spectral
density estimate divided by the sample size), and the $2.5\%$, $50\%$, and $
97.5\%$ credible regions are listed in table 1. The cross--correlation matrix
is presented in table 2. 

\section{Discussion}

The computational demands of Bayesian inference with large numbers of
parameters are best met with MCMC methods. These methods have demonstrated
their importance in numerous applications. As cosmological models grow in
complexity it will become necessary to use techniques such as those
discussed here in order to handle marginalization of parameters.

The demonstration of the Metropolis--Hastings\ routine with the toy model
was successful.  We plan to apply this technique to real CMB anisotropy power
spectrum measurements. With MCMC\ techniques one can easily extend this
method to \symbol{126}10 parameters. We are presently working on optimizing
the speed of the likelihood calculator for the larger parameter number.
Markov chain Monte Carlo methods provide a means of handling large parameter
numbers, and maintain a rigorous approach to Bayesian inference.  
They may prove to be essential for certain CMB parameter--determination
efforts in the near future.

\section{Acknowledgements}

This work was supported by the National Science Foundation, Royal Society of
New Zealand Marsden Fund, the University of Auckland Research
Committee, and Carleton College.  LK thanks the Institut d'Astrophysique
de Paris for their hospitality while some of this work was completed.

\bigskip {\LARGE Figure and Table Captions}

\bigskip

Figure 1. Trace and kernel density plots of the marginal posterior
distributions for the parameters $A$ (amp), $\omega _{b}$ (b), $\omega _{d}$
(d), and $n$ (tilt). The \textit{true} parameters are $A=100.0$, $\omega
_{b}=0.019$, $\omega _{d}=0.154$ and $n=1.0$.

\bigskip

Table 1. The posterior mean, standard deviation, time series standard error
(SE), and the $2.5\%$, $50\%$ (Median)\ and $97.5\%$ credible regions of the
parameters $A,$ $\omega _{b}$, $\omega _{d}$ and $n$ (cf. figure 1).

\bigskip

Table 2. The cross--correlation matrix of the parameters $A,$ $\omega _{b}$, $
\omega _{d}$ and $n$ (cf. figure 1).\bigskip 

\bigskip

\pagebreak \bigskip

\bigskip

TABLE\ 1

\begin{tabular}{|l|l|l|l|l|l|l|}
\hline
Parameter & Mean & SD & SE & $2.5\%$ & Median & $97.5\%$ \\ \hline
$A$ & $100.0$ & $0.729$ & $7.69x10^{-3}$ & $99.1$ & $100.0$ & $102.0$ \\ 
\hline
$\omega _{b}$ & $0.0187$ & $5.28x10^{-4}$ & $5.56x10^{-6}$ & $0.0177$ & $
0.0187$ & $0.0198$ \\ \hline
$\omega _{d}$ & $0.156$ & $2.27x10^{-3}$ & $2.39x10^{-5}$ & $0.151$ & $0.156$
& $0.160$ \\ \hline
$n$ & $0.971$ & $1.81x10^{-2}$ & $1.90x10^{-4}$ & $0.936$ & $0.971$ & $1.010$
\\ \hline
\end{tabular}

\bigskip

\bigskip

TABLE\ 2

\begin{tabular}{|l|l|l|l|l|}
\hline
Parameter & $A$ & $\omega _{b}$ & $\omega _{d}$ & $n$ \\ \hline
$A$ & 1.0 &  &  &  \\ \hline
$\omega _{b}$ & 0.220 & 1.0 &  &  \\ \hline
$\omega _{d}$ & 0.614 & 0.014 & 1.0 &  \\ \hline
$n$ & -0.189 & 0.420 & -0.440 & 1.0 \\ \hline
\end{tabular}

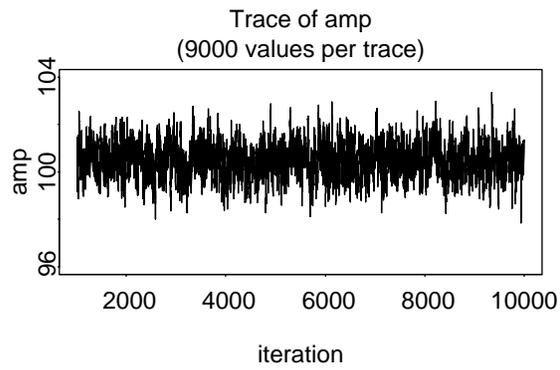
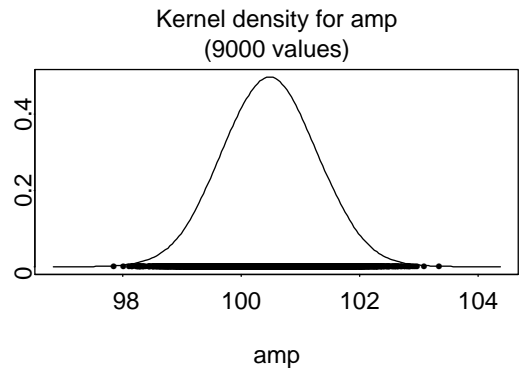
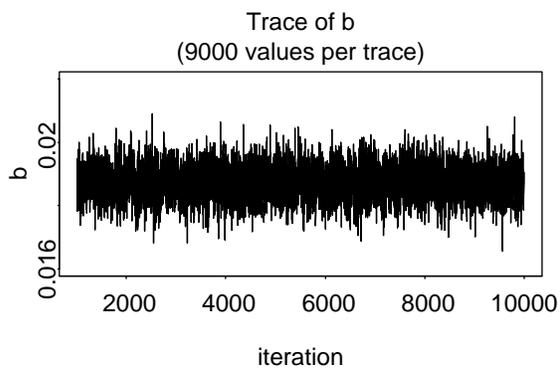
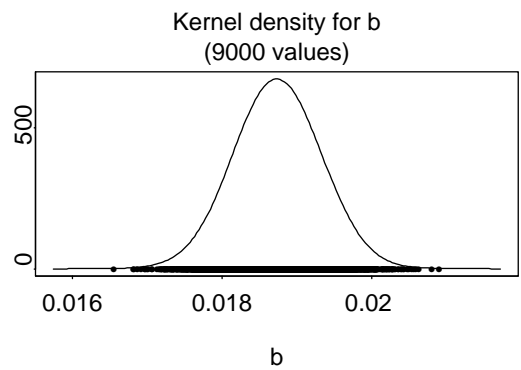
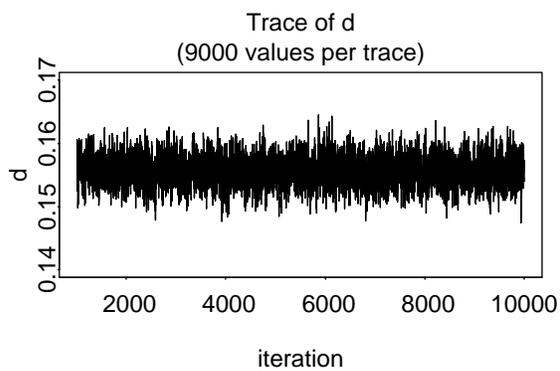
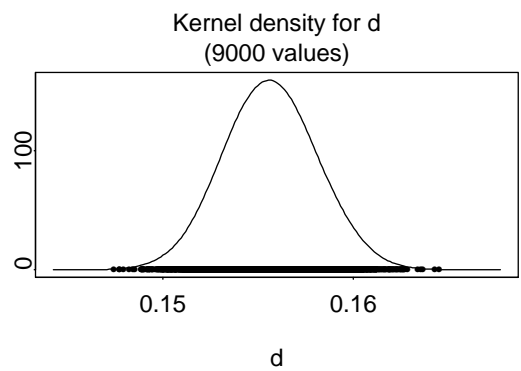
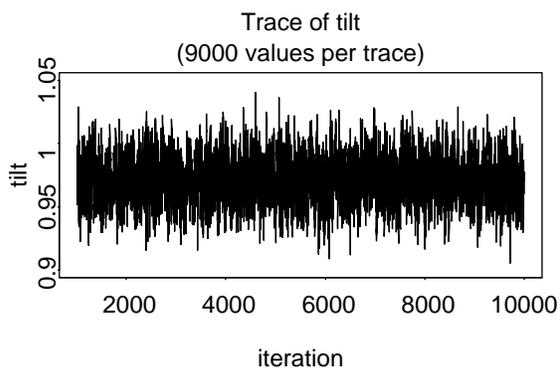
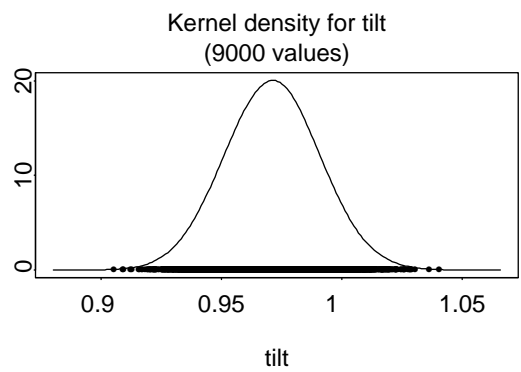
\end{document}